\documentclass[%
 aps,
rsi,%
 amsmath,amssymb,
 reprint,%
]{revtex4-1}

\usepackage{graphicx}% Include figure files
\usepackage{dcolumn}% Align table columns on decimal point
\usepackage{bm}% bold math
\usepackage{amsmath}               % great math stuff
\usepackage{amsfonts}              % for blackboard bold, etc
\usepackage{amsthm}                % better theorem environments
\usepackage{times}
\usepackage{comment}
\usepackage{array}
\usepackage{subfigure}
\usepackage{enumerate}
\usepackage{xcolor}
\begin{document}

\preprint{AIP/123-QED}

\title{Reactors as a source of antineutrinos:\\ 
effect of fuel loading and burnup for mixed oxide fuels }

\author{Adam Bernstein}
 \email{bernstein3@llnl.gov}
\author{Nathaniel S. Bowden}
\affiliation{Nuclear and Chemical Sciences Division, Lawrence Livermore National Laboratory}
\author{Anna S. Erickson}
\affiliation{770 State St NE, Nuclear and Radiological Engineering Program, Georgia Institute of Technology, Atlanta GA 30332}
 
\date{\today}% It is always \today, today,

\begin{abstract}

In a conventional light water reactor loaded with a range of uranium and plutonium-based fuel mixtures, the variation in antineutrino production over the cycle reflects both the initial core fissile inventory and its evolution. Under the assumption of constant thermal power, we calculate the rate at which antineutrinos are emitted from variously fueled cores, and the evolution of that rate as measured by a representative ton-scale antineutrino detector. We find that antineutrino flux decreases with burnup for Low Enriched Uranium cores, increases for full mixed-oxide (MOX) cores, and does not appreciably change for cores with a MOX fraction of approximately 75\%. Accounting for uncertainties in the fission yields, in the emitted antineutrino spectra, and the detector response function, we show that the difference in core-wide MOX fractions at least as small as 8\% can be distinguished using a hypothesis test. The test compares the evolution of the antineutrino rate relative to an initial value over part or all of the cycle. The use of relative rates reduces the sensitivity of the test to an independent thermal power measurement, making the result more robust against possible countermeasures. This rate-only approach also offers the potential advantage of reducing the cost and complexity of the antineutrino detectors used to verify the diversion, compared to methods that depend on the use of the antineutrino spectrum. A possible application is the verification of the disposition of surplus plutonium in nuclear reactors. 
 
\end{abstract}

\maketitle

\begin{comment}
Items to discuss:
- Need PACS codes
\end{comment}

\section{Introduction}
\label{Intro}
%\subsection{}
%\subsubsection{}

Nuclear reactors have been successfully used as a source of electron antineutrinos in numerous antineutrino physics experiments~\cite{Nikolo98, Boehm00, Eguchi03, Abe12} and have been proposed for use in safeguards 
applications~\cite{Klimov, Bernstein02, Bowden07, Bernstein08}. Reactors have played central role in characterizing antineutrino oscillations, due to their high intensity and reasonably well predicted antineutrino emission spectrum. In these experiments, one or more detectors are installed at fixed distances from a nuclear reactor core (or multiple reactors) to measure the antineutrino rate and, in some cases, the energy spectrum, through the inverse beta decay (IBD) process.  Antineutrino detectors have also been shown to be successful in non-intrusive tracking of the relative reactor power level \cite{Bernstein08} and burnup~\cite{nuburnup,Klimov}. 

In a typical light water reactor (LWR) on the order of 10$^{27}$ antineutrinos are emitted each day. A ton-scale detector placed $\sim25$~m from the reactor core can register as many as $\sim5000$ antineutrino events per day, providing information about the instantaneous relative reactor power and the gradual change in core-wide effective fuel burnup. The effective core-wide burnup, combined with a reactor physics simulation using known or declared operational parameters, can be used to deduce other parameters of the reactor operation, such as the burnup of individual fuel assemblies, or the operator-declared core loading. These derived quantities may be of interest for verification and monitoring tasks that require tracking the mass inventories of fissile materials at the assembly level.  

Antineutrino-based reactor monitoring appears promising for verification of in-core fuel isotopic composition and burnup as well as reactor operating history. As discussed in Bernstein et al.~\cite{Bernstein02}, key features of the methods include: 

\begin{enumerate}

\item {\bf Non-intrusiveness}: Antineutrino detectors are non-intrusive because they operate well outside the core, unconnected to any plant systems. 
\item {\bf Robustness against tampering}: This approach is resistant to spoofing, tampering or masking due to the penetrating power and great specificity of the antineutrino signature. 
%\item {\bf Cost}: Antineutrino detectors have been demonstrated to be cost-competitive  with the alternative calorimetric approach.
\item {\bf Continuity of knowledge}: Antineutrino detectors can provide continuous 
knowledge of the reactor operational status and thermal power levels.
\item {\bf Maturity level}: The method has been demonstrated with detectors
 at a wide range of nuclear reactors.
\end{enumerate}

The measured antineutrino  rate and spectrum depend not only on the reactor power level and initial fuel loading, but also on the subsequent evolution of the reactor core as fuel is irradiated during operation.  Changes in fission fraction distribution  as fuel burnup progresses also change the distribution of beta-decaying fission daughters. For this reason, antineutrino detectors can be used to track not only the relative reactor power, but also the core fissile composition.

One possible application of antineutrino monitoring is for non-intrusive monitoring of fuel loading and evolution in civil power reactors. In prior work, we and other authors have examined sensitivity to the evolution of Low Enriched Uranium (LEU) cores in Light Water Reactors (LWRs)~\cite{Bernstein02,Bula}. This represents the most typical configuration for civil power reactors worldwide. In this paper, we examine the case of LWRs fueled with a mixture of plutonium and uranium fuels, referred to as mixed-oxide (MOX) fuel. A reactor operating with LEU fuel will exhibit a different antineutrino rate and spectrum evolution in time compared to one loaded with MOX fuel. If the reactor core composition is known, changes in antineutrino flux can be used to determine operating power and core-wide fissile inventory. Together with core simulations, the spectral and rate information can also provide constraints on burnup and fissile inventories in individual assemblies.
%\textbf{weaken the language, reference to be provided above  - Adam }. 

The separated plutonium used to produce MOX fuel is categorized by the International Atomic Energy Agency (IAEA) as ``direct-use" material,  defined as ``nuclear material that can be used for the manufacture of nuclear explosive devices without transmutation or further enrichment"~\cite{IAEAdirect}. In this nomenclature, all other forms of nuclear material are referred to as ``indirect-use". A careful study of the variation in measured rates and spectra as a function of fuel loading is important because of the heightened sensitivity surrounding safeguards and monitoring of direct-use materials. For example, some LWRs in Europe and Japan already use MOX fuel~\cite{WNO}. In addition, countries such as the US, Russia and the UK are considering the use of LWRs (as well as other reactor types) to irradiate separated Weapons-Grade plutoniun (WGPu)~\cite{pmda,UKpmd} as a means of converting direct-use material into the indirect-use form of irradiated spent fuel, from which the plutonium is significantly more difficult to recover. The process of irradiating WGPu in reactor cores, as a means of rendering it more difficult to extract for weapons use, is often referred to as plutonium disposition. Both current commercial MOX-loaded cores, and any future plutonium disposition campaigns, may be able to take advantage of antineutrino-based safeguards, either to enhance existing safeguards protocols, or as part of a future verification regime specifically geared towards WGPu disposition.

In this paper, the antineutrino-based measurement approach is applied to a high-fidelity study of reactor core composition as a function of fuel burnup.  
In particular, we study the effects of loading a Westinghouse-type nuclear reactor core with various admixtures of low-enriched uranium  (LEU) and MOX fuel in realistic conditions.  
Extensive benchmarks have been performed for this reactor type to study the effects on reactor performance of conversion to partial MOX fuel loading~\cite{benchmark}. 
We performed full core simulations in order to account for variations in individual assembly fuel composition and burnup rates. The variation in neutron flux from assembly to assembly is an important parameter that is not captured in simpler one dimensional simulation approaches.

\section{\label{sec:Neut}Prediction of Antineutrino Flux}

Nuclear reactors are copious isotropic sources of antineutrinos, $\bar{\nu}_e$, 
because of the abundant fission reactions occurring in the core. A fission reaction is followed by production of two, or more rarely three, fission products which preferentially $\beta^-$ decay to achieve a stable neutron-proton ratio. On average, fission products will undergo three $\beta^-$ decays to reach stability, each emitting an antineutrino. The number of electron antineutrinos emitted varies with the number densities of the fission daughters and with the energy of the neutron inducing the fission reaction.  Hence, the composition of the nuclear core as well as the type of reactor play a role in the rate and spectrum of emitted antineutrinos.

\begin{comment}
A commonly used mechanism for detection of $\bar{\nu}_e$ is inverse beta decay:

 \begin{equation}
\bar{\nu}_e + p\to \beta^+ + n
\end{equation}
\\
\end{comment}
A commonly used mechanism for detection of $\bar{\nu}_e$ is the Inverse Beta Decay (IBD) reaction, which requires the antineutrinos to have a minimum energy of $1.804$~MeV. This threshold reduces the number of detectable  antineutrinos to  $\sim1.5-2$ per fission, depending on the isotope. IBD-based detectors are typically based on proton-rich organic materials and record the time and energy coincidence between the products of IBD, a prompt $\beta^+$ signal and neutron capture on hydrogen or other capture agent, e.g. Gd or $^6$Li, which is delayed by 30-200~$\mu$s. The energy deposited by the prompt $\beta^+$ is directly proportional to the antineutrino energy, permitting a measurement of the antineutrino energy spectrum.
 
In a typical LWR, four main isotopes are responsible for over 99\% of all fissions: $^{235}$U, $^{238}$U, $^{239}$Pu, and $^{241}$Pu. Extensive efforts in the antineutrino physics community have used measurements of the integral beta spectrum emitted by these isotopes to predict their antineutrino emissions. In the present study, we use a parameterization of  the antineutrino spectrum due to each fission isotope~\cite{huberPRC84} to calculate the expected changes in the summed antineutrino spectrum measured by the detector as the fuel evolves. 

\section{Comparison of LEU and MOX Fuel Characteristics}

To relate changes in antineutrino flux to initial core loading, we examine the evolution of the fuel in the core. Both the initial antineutrino rate and energy spectrum, and their subsequent evolution during the reactor operating cycle will vary depending on amount of MOX fuel that is loaded in the reactor core. Plutonium is both bred and destroyed in a nuclear reactor. Typical LWR reactors have plutonium conversion ratios of 0.6-0.7 for LEU and MOX assemblies indicating that plutonium production via a neutron capture on $^{238}$U in these reactors is insufficient to compensate for its destruction caused by fission. Fission is the main consumption mechanism for $^{239}$Pu. 

\begin{table} [b]
\caption{Comparison of LEU and MOX fuel.  }
\centering
\begin{ruledtabular}
\begin{tabular}{l c c}
\textbf{} & \textbf{UO$_2$ ($^{235}$U)} & \textbf{MOX ($^{239}$Pu)} \\
\hline
Thermal fission $\sigma$ (barn) & 577 & 741 \\
Delayed neutron fraction $\beta$ & 0.0065 & 0.0020 \\
%Neutron spectrum & softer & harder \\
%Burnup required to reach $^{240}$Pu/Pu = 20\% & 20 GWd/MTHM & 80 GWd/MTHM \\
Number of  $\bar{\nu}_e$\\ per fission ($E_{\bar{\nu}_e} >$ 1.806 MeV) & 1.92$\pm$0.2 &  1.45$\pm$0.2 \\
% is this number of nubars correct for the fuel composition ? It looks like it is just the number emitted by Pu and U, rather than that MOX and LEU
Energy released per fission (MeV) & 201.7$\pm$0.6 &  210.0$\pm$0.9 \\
\end{tabular}
\end{ruledtabular}
\label{tab:table1}
\end{table}

Table \ref{tab:table1} compares typical LEU and MOX fuel characteristics. The thermal  fission cross section of $^{239}$Pu is higher than that for $^{235}$U which results in an abrupt thermal flux gradient and power peaking changes at the interface of LEU and MOX assemblies. 
The localized power peaking caused by this high cross section imposes MOX fuel assembly design and location constraints beyond those required for LEU fuel. 
In particular, MOX fuel often features two or three zones of different plutonium enrichment to smooth the thermal flux gradient within the assembly as well as at the assembly interface.  
The delayed neutron fraction, $\beta$, of  $^{239}$Pu is much lower than for $^{235}$U and as a result MOX fuel poses greater reactor control and safety challenges. 
This effect is more pronounced for WGPu fuels because of their higher fraction of $^{239}$Pu.  
Another important distinction between LEU and MOX fuel in a thermal reactor is the fission neutron spectrum. 
The harder neutron flux characteristic of MOX fuel significantly reduces the utility of 
chemical shims and control rods. 
The combination of lower delayed neutron fraction and harder neutron spectrum in MOX fuel makes it difficult to achieve a 100\% loading in a typical PWR core. 

Averaged over energy, the number of antineutrinos released per fission with energies above the inverse beta threshold is $\sim 25$\% lower for MOX fuels than for LEU. Thus, the expected antineutrino flux from a MOX-loaded core will be significantly lower than that of an LEU core, at least in the beginning of the cycle. In addition, MOX fuels generate more heat per fission than LEU fuels. As a result, fewer fissions are required to achieve the same core power, further reducing the total number of antineutrinos emitted. 

Operation of a full MOX core in a conventional LWR would require an extensive evaluation, and may not be possible for many reactors without significant modifications to reactor systems. However, various designs offering full-core MOX loading do exist, for example the CE System-80~\cite{study94} and AP-600/1000 reactor designs. The advantage of having a 100\% MOX loaded core is not only higher plutonium throughput, but also an elimination of the severe flux gradients and power peaking effects characteristic of partially-loaded cores.  

%Full MOX core options have not been extensively  
%researched, but have the potential to play an important role in plutonium 
%disposition campaigns.  

% \section{Analysis of Antineutrino Rates Using MCNPX}
 \section{Prediction of Fission Rates Using MCNPX}

An assembly-level analysis of LEU and MOX-loaded cores was performed with MCNPX2.7 coupled with CINDER90. MCNPX~\cite{MCNPX} calculates the steady state reaction rates, flux, and normalization parameters, and CINDER90 provides time-dependent isotope buildup and depletion.  MCNPX is a general purpose Monte Carlo code capable of modeling complex geometries.  It uses cross-section data evaluated from ENDF/B-VII. 

One eighth of the core was modeled because the symmetry in core loading  pattern allows for reflective boundary conditions. 17 unique assemblies  were defined in the model based on isotopic composition. In this analysis, burnable absorbers integrated into the fuel assemblies were modeled in full detail.  Control rods and chemical shim reactivity controls were not included.  CINDER90 uses a solver for independent contributions to atom densities 
and is capable of tracking concentrations for  3456 isotopes (1325 fission products). Burnup calculations were performed  with MCNPX Tier 3 fission product setting, which allows tracking of all available isotopes. The time-dependent fission rates output by the MCNPX/CINDER90 package, are convolved with parameterized ~\cite{huberPRC84} antineutrino spectral densities (defined as the number of antineutrinos per MeV and fission), to obtain the emitted flux of antineutrinos as a function of energy and time.   

\section{\label{sec:LWR}Study of Westinghouse-type LWR core}
\subsection{Nominal Core Configuration}

This study is based on a characteristic PWR core, one of the most common reactor types in the US and worldwide.
The core geometry is modeled on that of a standard Westinghouse-type reactor, with core height of 4 m and  thermal power of 3200~MW. Because the physics of MOX assemblies is very different from LEU assemblies, and because the interface between the two types of  assemblies plays a significant role in burnup accumulation and distribution, we modeled individual assemblies rather than a homogenized core. The geometry and composition parameters were adopted from an OECD Nuclear Energy Agency Nuclear Science Committee benchmark~\cite{benchmark}.
The relative complexity of the fuel assembly arrangement discussed below is due to the non-trivial neutron physics that results from a partial MOX core. 

\begin{figure}[b]
\setlength{\abovecaptionskip}{-5pt}
\begin{center}
\includegraphics[width=0.48\textwidth]{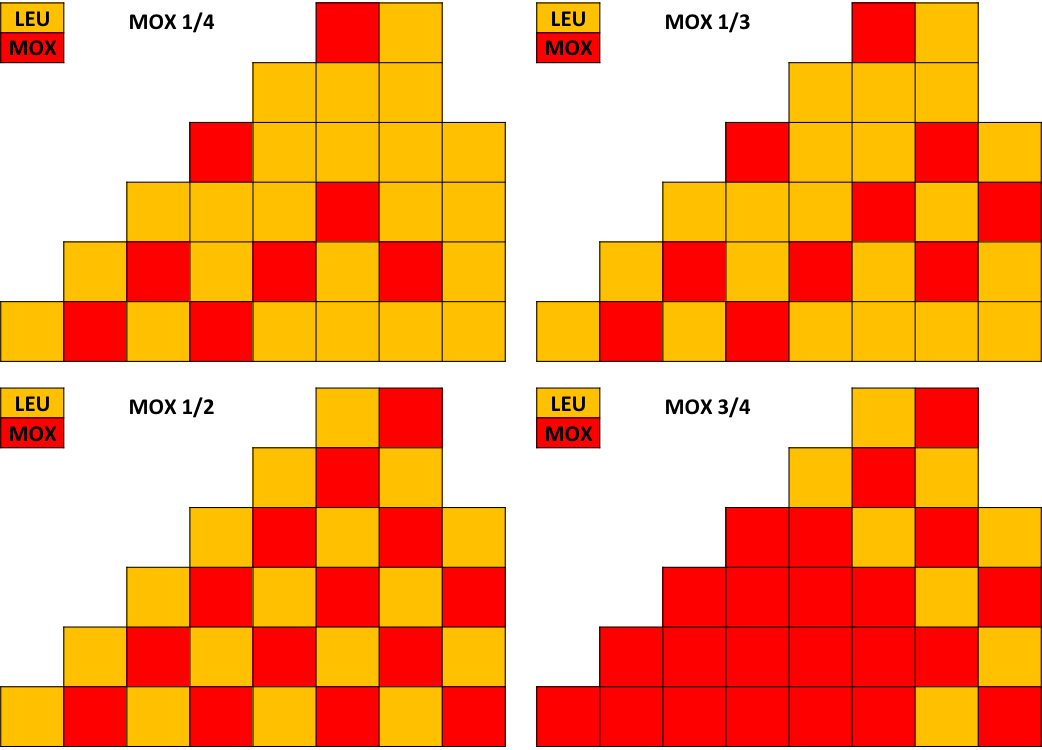}
\end{center}
\caption{Illustration of core layout for several MOX loading fractions.}
\label{fig:moxFractions}
\end{figure}

MOX assemblies are designed to have geometric, power, and neutronic compatibility 
with existing LEU assemblies in a core. The initial simulated core geometry 
and isotopic composition correspond to a simplified single-batch 
(no once and twice burned assemblies), single-enrichment core.  The antineutrino rate emitted by a core fully or partially loaded with MOX fuel primarily depends on two factors: the fraction of the core that is MOX and core-wide fuel burnup. To understand the effect of both factors, the LWR core was loaded with fresh LEU and weapons-grade MOX (WGMOX) assemblies, and irradiated without assembly shuffling for three cycles, to a burnup of about 60 MWd/kgHM. Plutonium is considered weapon-grade when the ratio of $^{240}$Pu to $^{239}$Pu is less than 7\%. Figure~\ref{fig:moxFractions} depicts the MOX loading fractions in the core that were considered in the analysis. Core reactivity in each case was similar to the nominal 100\%  MOX case, in order to avoid comparison of fundamentally different scenarios. 

\subsection{Description of the LEU and MOX Assemblies Models}

\begin{figure}[h!]
\begin{center}
\includegraphics[width=0.28\textwidth]{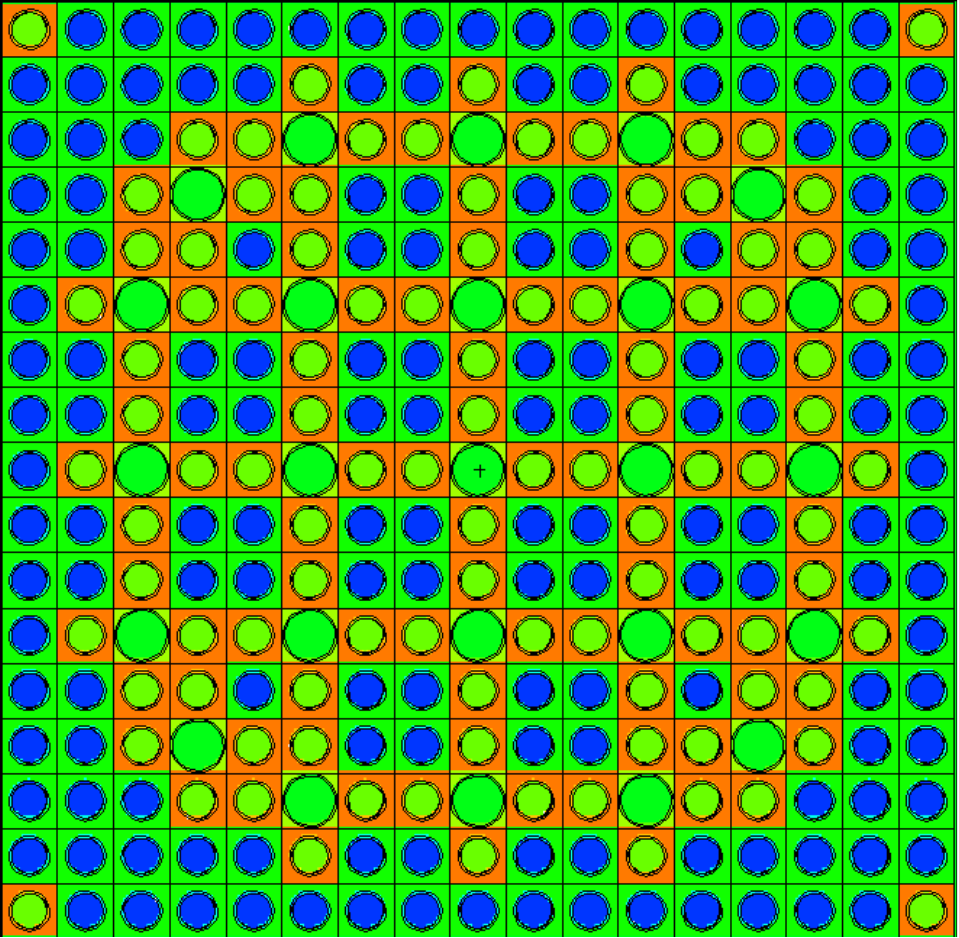}
\end{center}
\caption{MCNP model of a UO$_2$ assembly. Blue fuel pins correspond to regular 
pins, and green/orange fuel pins contain IFBA. The assembly also includes 
25 guide tubes.} 
\label{fig:LEU}
\end{figure}

Fuel composition and geometric parameters of MOX and LEU assemblies 
are provided in Table~\ref{tab:table3}. Note that LEU fuel assemblies have two enrichment values: 
4.2\% and 4.5\% of $^{235}$U. About $1/3$ of the fuel pins in an LEU assembly are
 Integral Fuel Burnable Absorber (IFBA) pins, as shown in Figure~\ref{fig:LEU}. 
 IFBA is a thin outer layer of ZrB$_2$ on fuel pellets and is consumed with 
 irradiation.  IFBA pins are heterogeneously dispersed throughout the assembly 
 to lower the reactivity of fresh fuel and to flatten the peaking factors in the core. 

\begin{table} [t!]
\caption{Fuel composition used in the analysis.}
\begin{center}
\setlength{\extrarowheight}{1.75pt}
\begin{tabular}{ >{\centering\arraybackslash}m{2cm}  
                             >{\centering\arraybackslash}m{2cm}  
                             >{\centering\arraybackslash}m{2cm} 
                             >{\centering\arraybackslash}m{2cm} }
 \hline
  \multicolumn{4}{c}{Uranium vector (wt\%)} \\
 \hline
$^{234}$U   & $^{235}$U & $^{236}$U & $^{238}$U \\
0.002  & 0.2 & 0.001 & 99.797 \\
 \hline
  \multicolumn{4}{c}{WG plutonium vector (wt\%)} \\
 \hline
$^{239}$Pu   & $^{240}$Pu &  $^{241}$Pu &  $^{242}$Pu \\
93.6 & 5.9 & 0.4 & 0.1 \\\hline
 \multicolumn{4}{c}{RG plutonium vector, 33 MWd/kgHM burnup (wt\%)} \\
 \hline
$^{239}$Pu   & $^{240}$Pu &  $^{241}$Pu &  $^{242}$Pu \\
56.0& 24.1 & 12.8 & 5.4 \\\hline
\end{tabular}
\end{center}
\label{tab:table3}
\end{table}

\begin{figure}[h!]
\begin{center}
\includegraphics[width=0.48\textwidth]{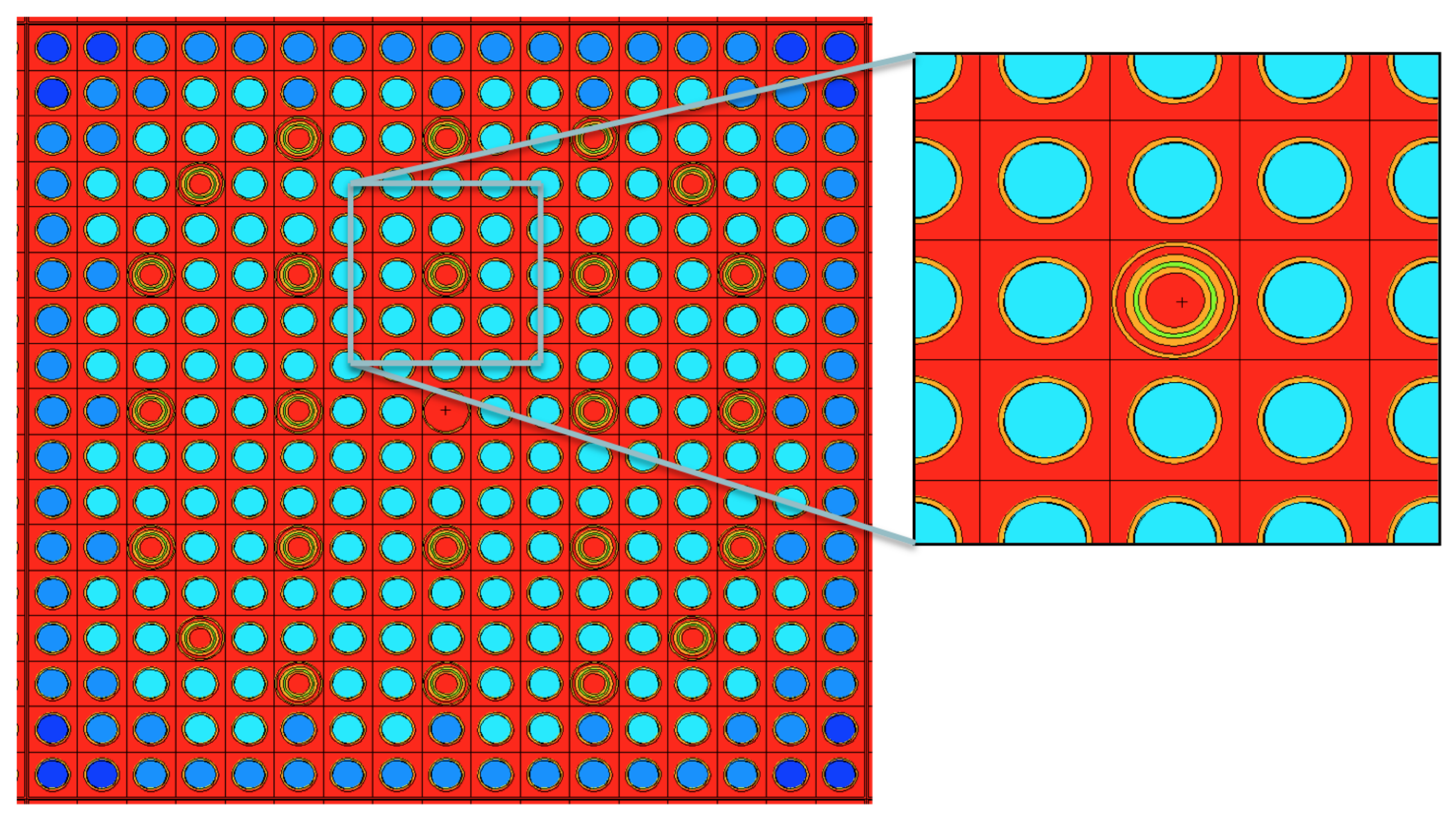}
\end{center}
\caption{MCNP model of a MOX assembly. Shades of blue represent fuel pins 
with three plutonium enrichments. Insert image shows details of WABA pins in the 
assembly. The central tube is an instrumentation tube.}
\label{fig:MOX}
\end{figure}

The loading pattern of MOX assemblies is heterogeneous and includes three 
enrichment zones in each assembly.  The enrichment varies from 2.5\% to 5.0\% 
of $^{239}$Pu, with lower enrichment near the edges of the assembly and higher 
enrichments near the center. This configuration helps to reduce the power 
peaking characteristic of LEU-MOX interface.  The matrix used in MOX fuel 
was depleted uranium with 0.2 wt\% of $^{235}$U. In MOX assemblies, the space 
normally occupied by the control rod positions is reserved for 24 Wet Annular 
Burnable Absorber (WABA) pins. The three-zone enrichment and WABA distribution 
is shown in Figure~\ref{fig:MOX}. WABA pins have a complex annular structure 
illustrated in an insert of Figure~\ref{fig:MOX}. The absorber used in WABA is 
Al$_2$O$_3$-B$_4$C. Multiple WABA pins are required in MOX assemblies 
because of their harder neutron spectrum.   

\section{\label{sec:MOXAR}Relation between MOX Loading and Antineutrino Rate}

% \begin{figure}[t!]
% %\setlength{\abovecaptionskip}{-15pt}
% %\setlength{\belowcaptionskip}{0pt}
% \begin{center}
% \subfigure[~Fission fraction of $^{235}$U] {
% \includegraphics[width=0.48\textwidth]{./Figures/FR_u235}
% \label{fig:FR_u235} }
% \subfigure[~Fission fraction of $^{239}$Pu] {
% \includegraphics[width=0.48\textwidth]{./Figures/FR_Pu239}
% \label{fig:FR_Pu239}}
% \subfigure[~Fission fraction of $^{241}$Pu] {
% \includegraphics[width=0.48\textwidth]{./Figures/FR_Pu241}
% \label{fig:FR_Pu241}}
% \end{center}
% \setlength{\abovecaptionskip}{-5pt}
% %\setlength{\belowcaptionskip}{0pt}
% \caption{Change in fission fractions of $^{235}$U, $^{239}$Pu, and $^{241}$Pu 
% as a function of core burnup and MOX loading. Note that the burnup axis orientations in 
% $^{239}$Pu plot is reversed for clarity.}
% \label{fig:fissFractions}
% \end{figure}

\begin{figure}[t!]
\begin{center}
%\subfigure[~mox000] {
\includegraphics[width=0.44\textwidth]{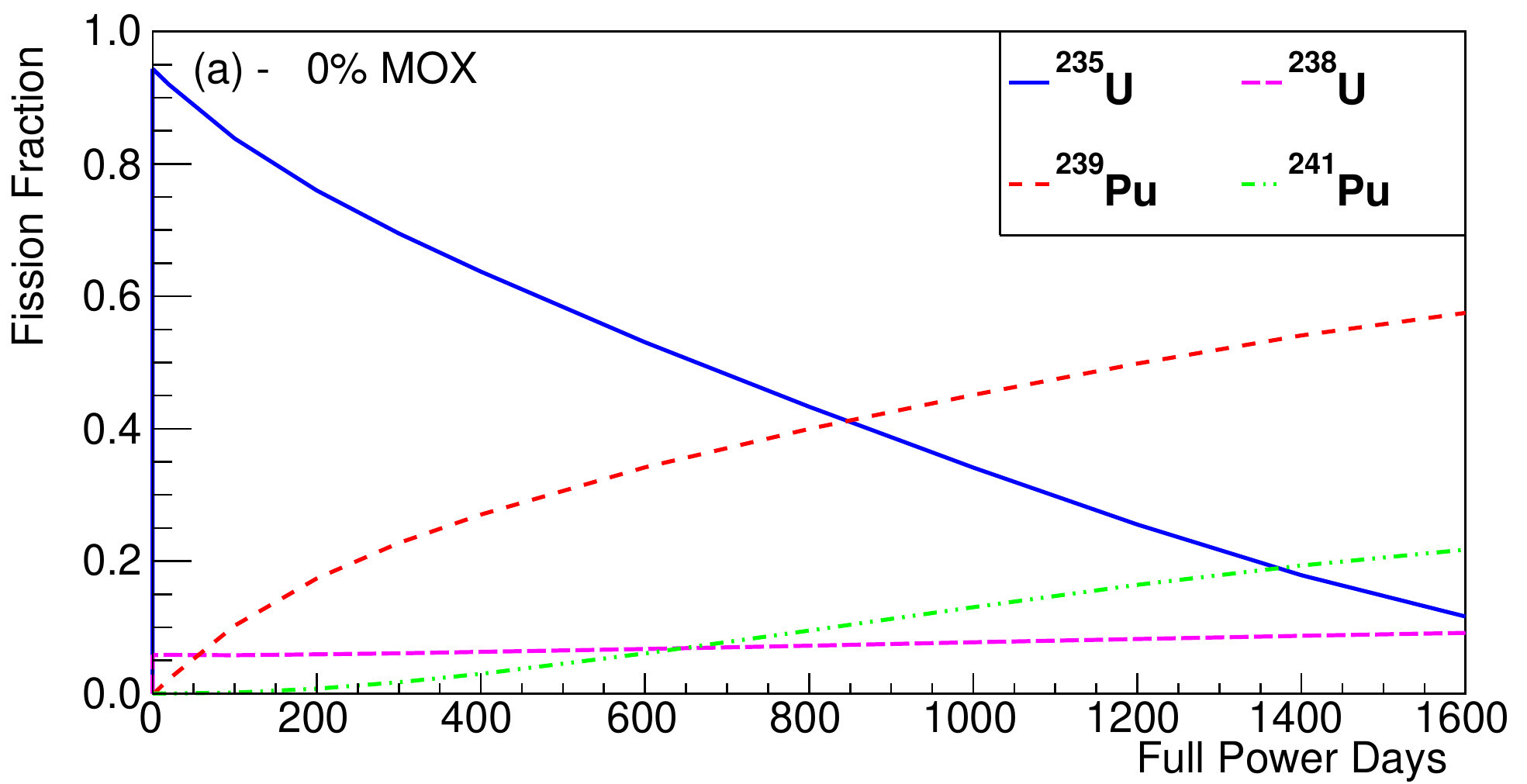}
%\label{fig:FR_mox000} }
%\subfigure[~mox025] {
\includegraphics[width=0.44\textwidth]{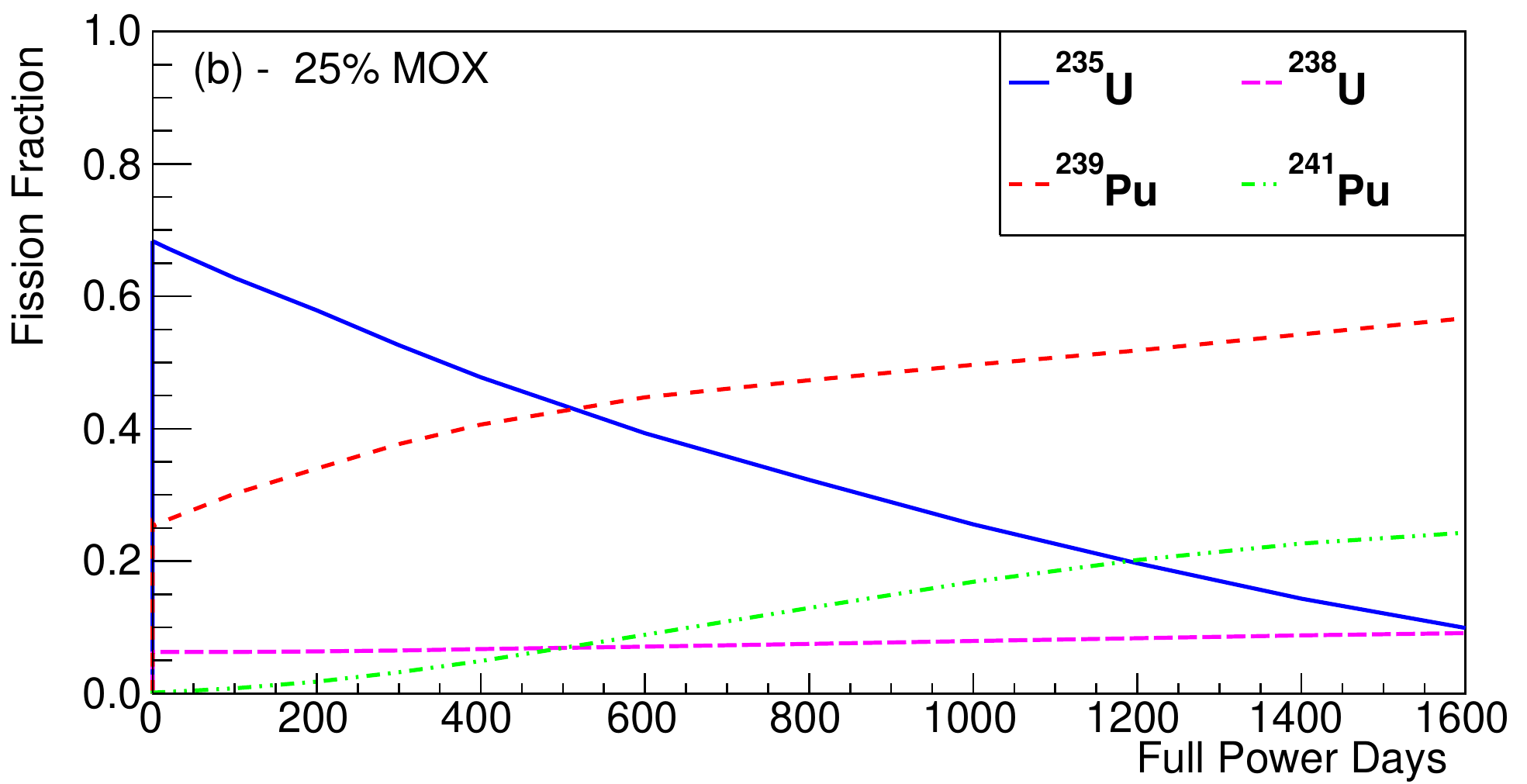}
%\label{fig:FR_mox025}}
%\subfigure[~mox033] {
%\includegraphics[width=0.40\textwidth]{./Figures/mox033}
%\label{fig:FR_mox033}}
%\subfigure[~mox059] {
\includegraphics[width=0.44\textwidth]{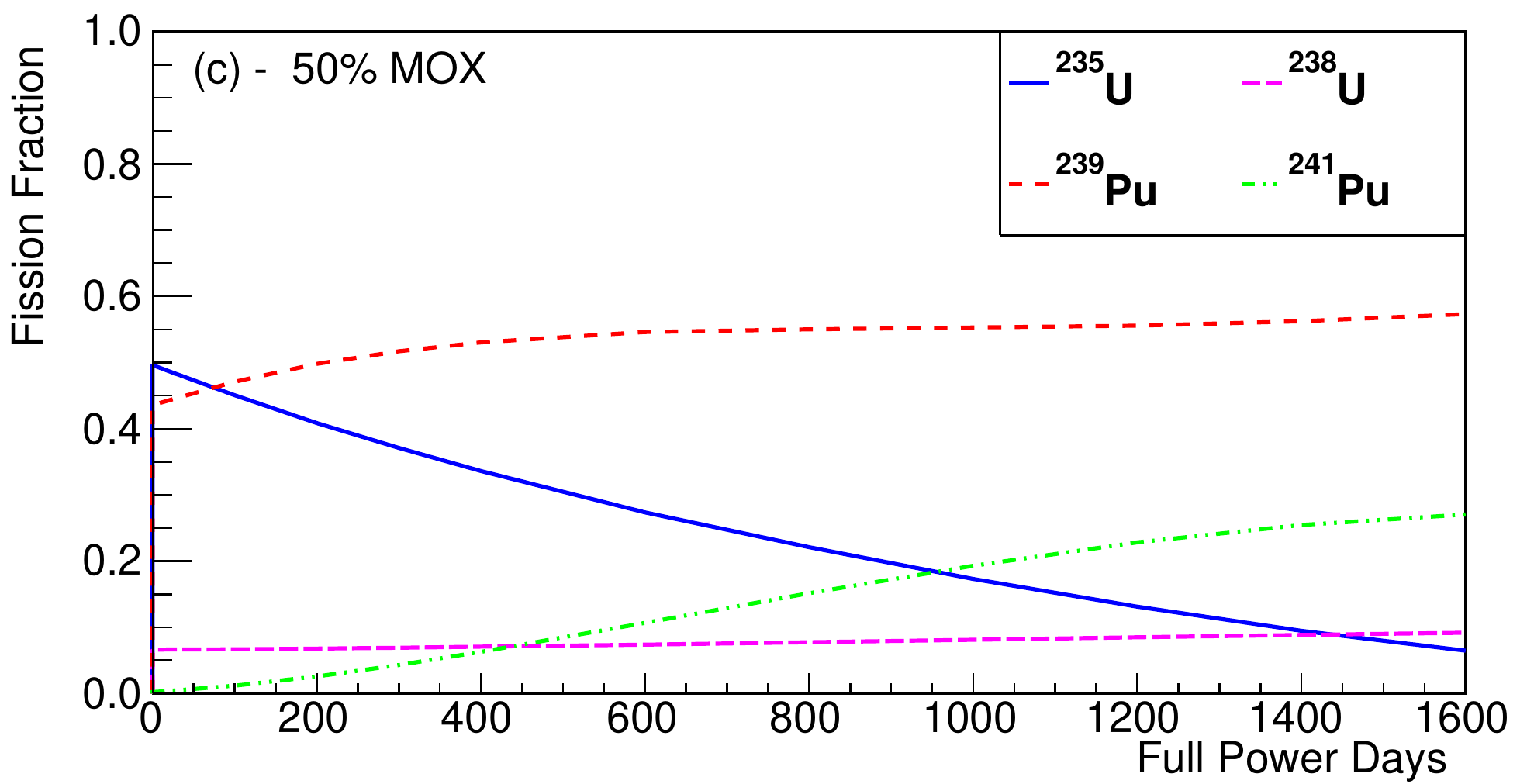}
%\label{fig:FR_mox075}}
%\subfigure[~mox033] {
\includegraphics[width=0.44\textwidth]{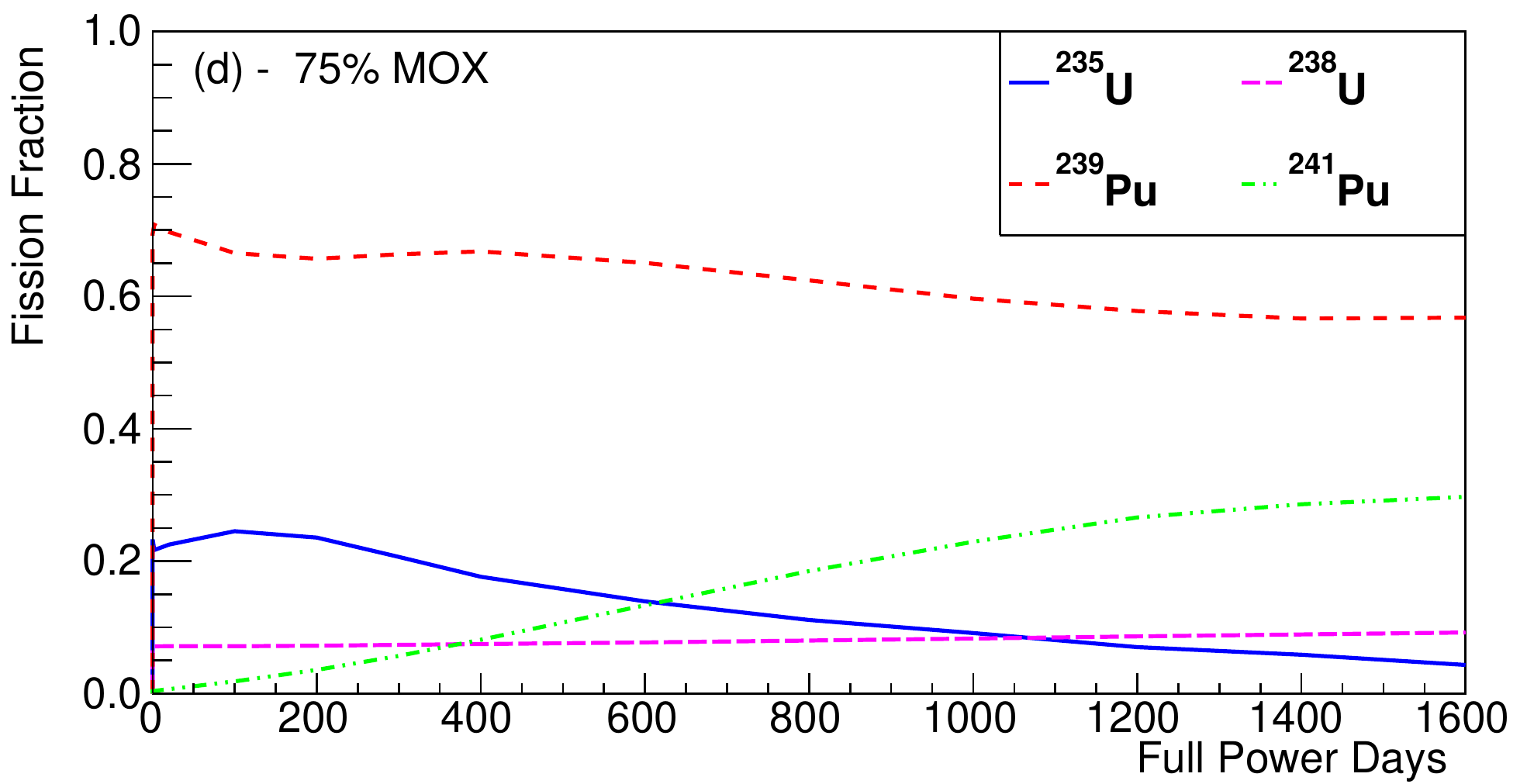}
%\label{fig:FR_mox100}}
%\subfigure[~mox100] {
\includegraphics[width=0.44\textwidth]{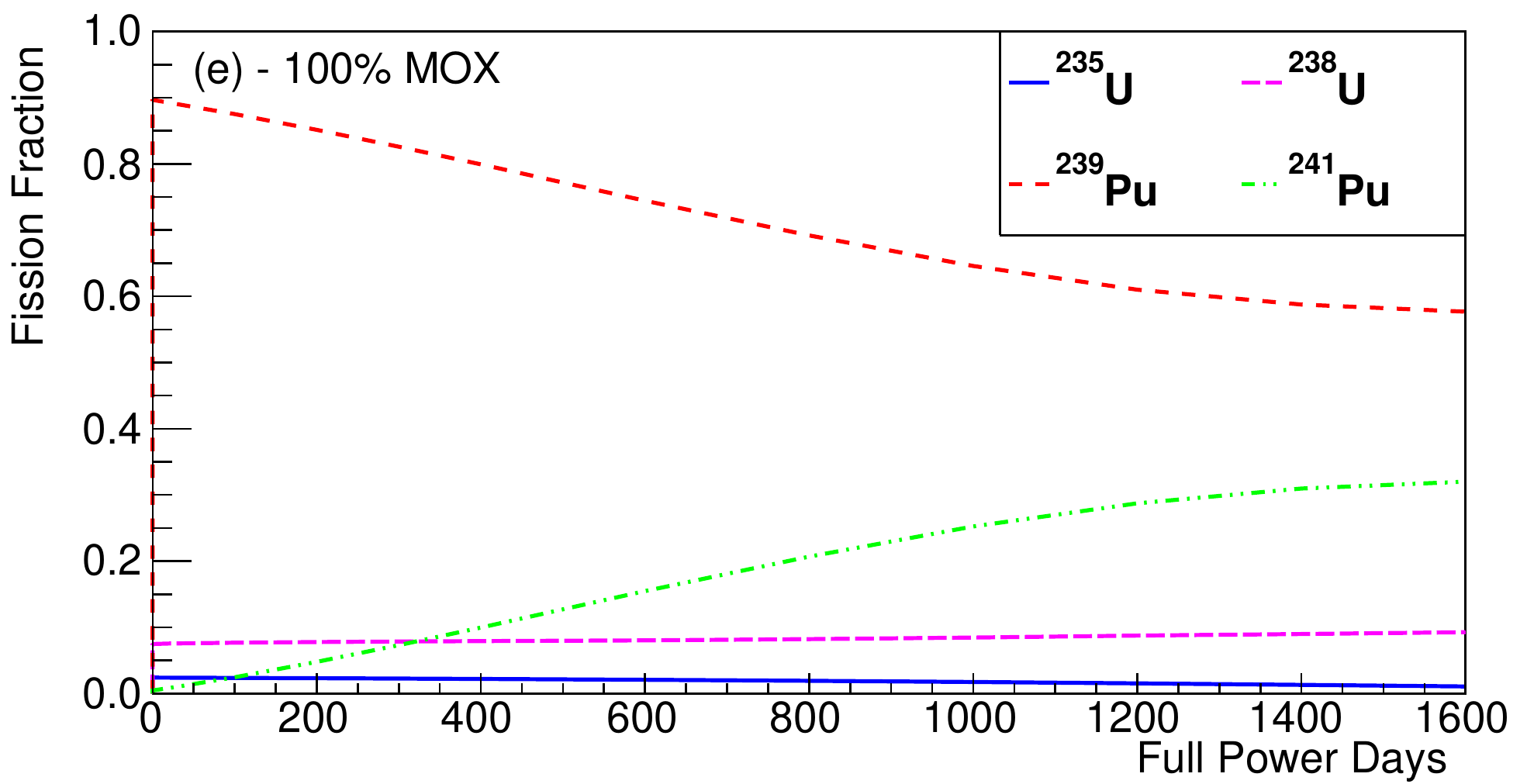}
%\label{fig:FR_mox033}}
\end{center}
\setlength{\abovecaptionskip}{-5pt}
\caption{Change in fission fractions of $^{235}$U, $^{238}$U, $^{239}$Pu, and $^{241}$Pu 
as a function of cycle time (core burnup) and initial MOX loading. Loadings considered are (a) full LEU, (b) 25\% MOX, (c) 50\% MOX, (d) 75\% MOX, and (e) full MOX.  Note relatively small effect of the MOX fraction on fission rates of $^{238}$U and $^{241}$Pu isotopes. 
%Note that the burnup axis orientations in $^{239}$Pu plot is reversed for clarity.
}
\label{fig:fissFractions}
\end{figure}

To investigate the initial antineutrino emission rate as well as its time-dependent evolution, the pure LEU core was compared to a 100\% MOX core and with fractional MOX cores, loaded as illustrated in Figure~\ref{fig:moxFractions}. 
% I changed the wording because there are additional fractions in Figure 6. 
Figure~\ref{fig:fissFractions} shows the evolution of the fission fractions for $^{235}$U, $^{239}$Pu, and $^{241}$Pu, as a function of the core burnup and MOX loading.
The contribution of $^{238}$U to the total fission rate is generally  below 10\% and does not change considerably for any initial loading considered here. The fission fraction of $^{235}$U decreases linearly with MOX loading and is also inversely proportional to burnup. The fission rate of $^{241}$Pu is a good indicator of burnup regardless of the core MOX fraction since the initial fraction of $^{241}$Pu in the fuel is low for WGPu fuel. 
On the other hand, the behavior of  $^{239}$Pu is highly dependent on both the burnup and the MOX fraction. As the MOX loading increases, the fission  rate of $^{239}$Pu increases nearly linearly for fuel of any burnup.   

\begin{figure}[ht]
\begin{center}
\includegraphics[width=0.48\textwidth]{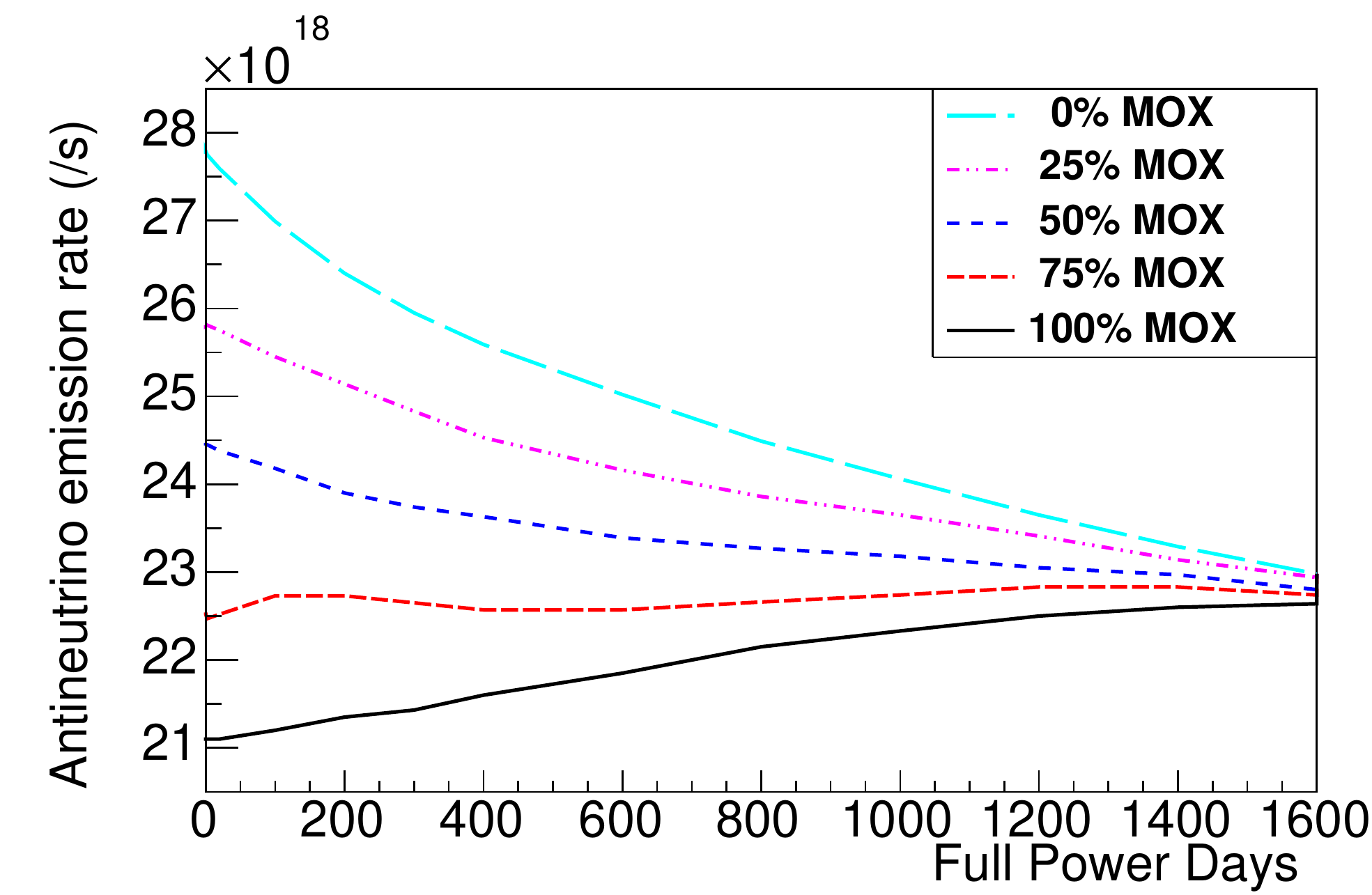}
\label{fig:neutRateAllCases} 
\end{center}
\setlength{\belowcaptionskip}{0pt}
\caption{ The total emitted antineutrino rate above the IBD threshold as a function of burnup. Note that when the core contains 75\% of MOX, antineutrino emission does not change  appreciably with burnup. 0\% MOX corresponds to a traditional LEU-type core.}
\label{fig:neutFractions-2}
\end{figure}

The combined evolution of all fissionable isotopes is reflected in the emitted antineutrino flux. The emitted flux is estimated by multiplying the simulated fission rates for each isotope and time step with the corresponding isotopic antineutrino spectral density and summing over all isotopes.  Figure~\ref{fig:neutFractions-2} shows the change in the emitted antineutrino flux above IBD threshold as function of MOX loading in the core and fuel irradiation time. 
%CHECK \textbf{above IBD threshold} as function of MOX loading in the core and fuel irradiation time. 

ntineutrino rate detected from a reactor core with a particular initial fuel loading can be expressed as:

\begin{equation}
\label{eq:Eq1}
\frac{\text{d}N_{\bar{\nu}_e}}{\text{d}t}=\gamma\left(1+n(t) \right)P_\text{th}
\end{equation}

In this equation, $\frac{\text{d}N_{\bar{\nu}_e}}{\text{d}t}$ is the detected antineutrino count rate, $\gamma$ is a constant that incorporates the effect of detector size, position, and efficiency, $n(t)$ is a time dependent sum over the fission rates of each isotope, and $P_\text{th}$ is the reactor core thermal power. 
If the core power is known during the cycle, then fissile content at each moment in the reactor core evolution can be inferred.

The change in antineutrino rate relative to the beginning-of-cycle 
rate is proportional to the amount of plutonium in the core. Thus, the ratio of the instantaneous rate to the initial rate throughout the cycle can indicate the MOX fraction in the core as long as the power is fixed. On the other hand, if the core MOX loading is known and constant throughout the cycle, reactor power can be 
 determined. This reflects a fundamental advantage of antineutrino monitors over  conventional power monitors, which can determine the total core power, but are unable to quantify plutonium in the core. Note that for an LEU core, antineutrino emission decreases with burnup, while for a full MOX core, the antineutrino detection rate increases with burnup. For cores with about 75\% loading of MOX, antineutrino emission does not change  appreciably with burnup. 
%This change in antineutrino emission rate due to  fuel burnup has been experimentally demonstrated for light water reactors~\cite{Bowden07} . Figure~\ref{fig:neutFractions-2} shows the change in the absolute detectable (for energies above 1.8~MeV) antineutrino rate as a function of fuel burnup. 
The absolute antineutrino rate depends on the type of the fuel loaded in the core and conveys  the information about fuel burnup in a unique manner for any MOX fraction.
Figure~\ref{fig:neutFractions-2} reflects the information carried by Eq.~\ref{eq:Eq1}, with (1+$n(t)$) describing the shape.

\section{Comparison of the detected signal from WGMOX, RGMOX, and LEU}
In Section~\ref{sec:MOXAR}, we showed that the evolution of the emitted antineutrino rate over a cycle is a function of the MOX loading in the core. In this section, we elaborate on this basic result by examining the detected rates with a realistic detector configuration. We also compare the rate evolution for Reactor Grade Plutonium (RGPu) to those for LEU and WGPu MOX loadings.   %(the Pu isotopic vector is provided in Table~\ref{tab:table3}) 
The presence of RGPu in the core may be distinguished because the additional $^{241}$Pu has an emission spectrum that differs from that of $^{239}$Pu, affecting the detected antineutrino rate evolution. 

\subsection{Comparison of Detected Antineutrino Rates}
In order to estimate detected antineutrino rates, we assume a 5 metric ton liquid scintillator detector, at 25 meter standoff from the reactor core center. Backgrounds are taken to be 25\% of the antineutrino rate averaged over all evolutions. This background ratio is consistent with that achieved in a previous reactor monitoring demonstration~\cite{Bowden07}.

The highest-performing modern antineutrino detectors are able to set a detection threshold at the inverse beta limit of 1.8 MeV for the antineutrino energy (e.g.~\cite{Abe12}). In some deployments, a higher threshold may be desirable in order to suppress backgrounds. We varied the detector energy threshold from $1.8$~MeV to $3.3$~MeV to determine the effect of this parameter on sensitivity to changes in core loading.  We assume detection efficiencies of 40\% ($1.8$~MeV  threshold) or 30\% ($3.3$~MeV threshold). The total rate of detected antineutrinos drops by a factor of about 1.5 as the threshold is increased. 

The estimated  efficiency is based on reasonable extrapolations from previous shallow-depth monitoring detector designs~\cite{Bula}.  In the detector, these thresholds would correspond to energy depositions of $\sim 1$~MeV and $\sim 2.5$~MeV, where the latter value is close to the highest energy naturally occurring background gamma-ray.

In this section and the following, all plots show the results assuming a $1.8$~MeV threshold. As shown in Table \ref{RSTtab}, section~\ref{sec:hyptest}, our principal results differ only modestly with the higher detection threshold imposed. 

Using the core powers and fuel loadings defined earlier, we can estimate rates for each fuel loading and at each burnup step in our MCNPX simulation. 

To determine whether the fuel loading effects can be distinguished in realistic deployments, we must account for systematic and statistical uncertainties. 
Table \ref{tab:sysratetab} shows the assumed sources of systematic error and their magnitudes, expressed as a percentage of the detected rate. Systematic uncertainties on the fission fractions of each isotope are taken from an analysis of the Takahama benchmark~\cite{Takahama}. At 11\%, $^{241}$Pu has the highest uncertainty in fission fraction, partially offset by its relatively small contribution to the total fission rate.  Systematic uncertainties in the emitted antineutrino energy spectra errors are taken from the Huber parametrization~\cite{huberPRC84}. Correlated and uncorrelated spectral errors are treated separately in the uncertainty analysis. For the correlated errors, a one sigma (fully correlated) shift is made for all spectral bins for each isotope. Uncorrelated errors are applied randomly bin by bin, assuming Gaussian-distributed errors. The relatively small systematic uncertainty on the total antineutrino output from each isotope is estimated by comparing the estimated output from the Huber~\cite{huberPRC84} and Fallot~\cite{Fallot} flux parametrizations. Statistical errors are determined as the Gaussian error on the daily number of counts at each burnup step, including the statistical error associated with subtraction of the 25\% background acquired over a 30-day period.  

\begin{table}
\protect\caption{Summary of rate estimate systematic uncertainties  \label{tab:sysratetab}}
\begin{ruledtabular}
\begin{tabular}{lccc}
Source of Uncertainty & Magnitude \\
\hline 
Fission fraction & $\pm 2-11 $\%   \\
uncorrelated spectral errors & $\pm 2-7$\%  \\
correlated spectral errors & $\pm 2-7$\%  \\
Total antineutrinos per fission for each isotope & $ <1\% $ \\
Detector upper/lower energy threshold & $ <1\% $  \\
\end{tabular}
\end{ruledtabular}
\end{table}

\begin{figure}[t]
\begin{center}
 \includegraphics[width=0.48\textwidth]{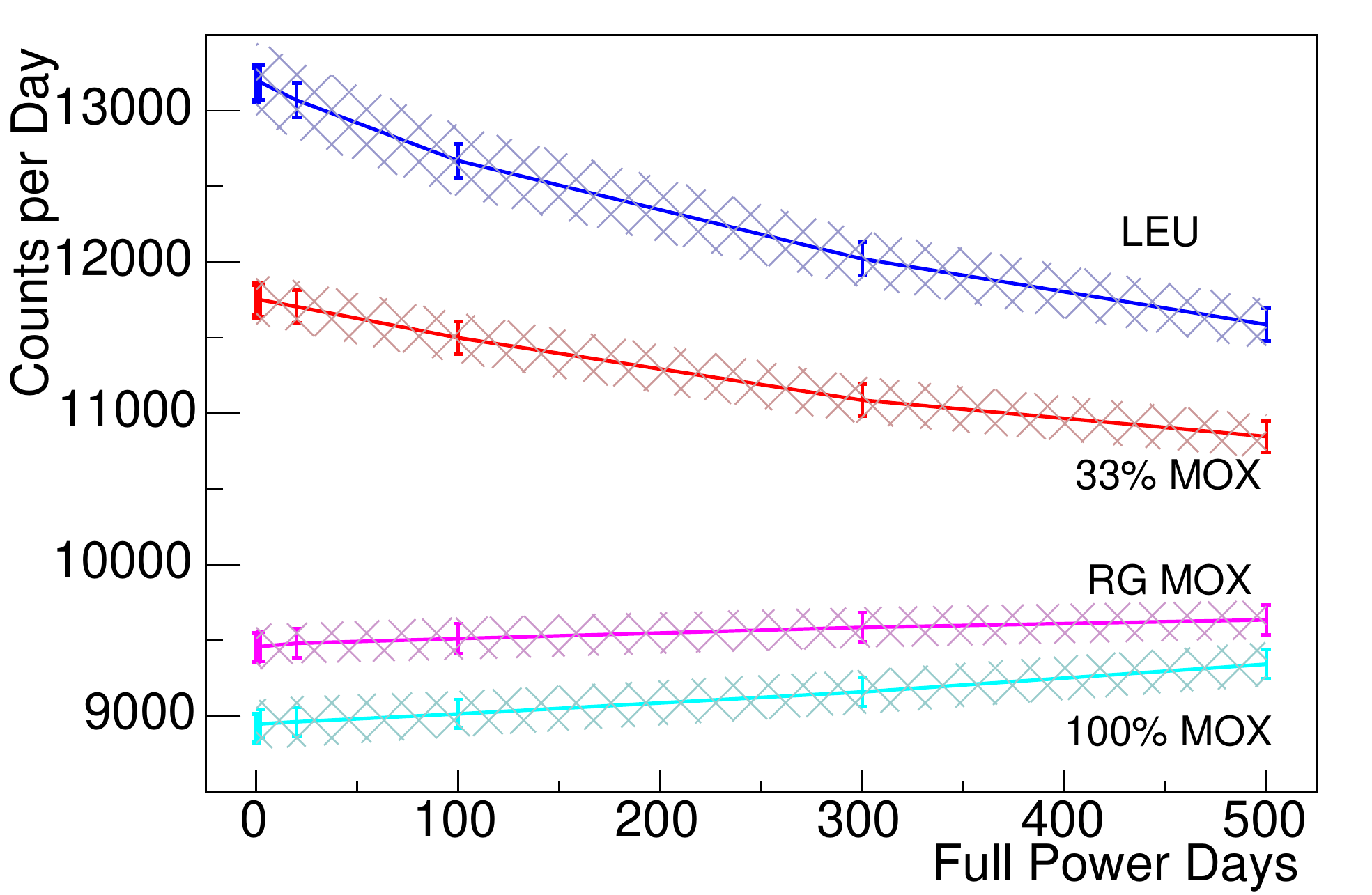}
% \label{fig:sigAbs1MeV}} 
 \end{center}
 \setlength{\belowcaptionskip}{0pt}
 \caption{The detected antineutrino rate evolution,  for four reactor fuel loadings. Systematic (hashed area) and statistical uncertainties are given for 24 hour integration windows.}% \textbf{check & error order threshold?} }
 \label{fig:neutAbs}
 \end{figure}

Figure~\ref{fig:neutAbs} shows the resulting detected antineutrino rates with estimated systematic and statistical uncertainties for WGPu and RGPu fuels, an LEU core and a partially loaded (1/3) WGPu MOX core. 
As seen in the figure, the total detected rate from any of  MOX-loaded core is always less than that of a full LEU core.

\begin{figure}[ht!]
\begin{center}
%\subfigure[
% The detected antineutrino rate, normalized to unity at zero burnup (Day 0 of the cycle)
%] {
\includegraphics[width=0.48\textwidth]{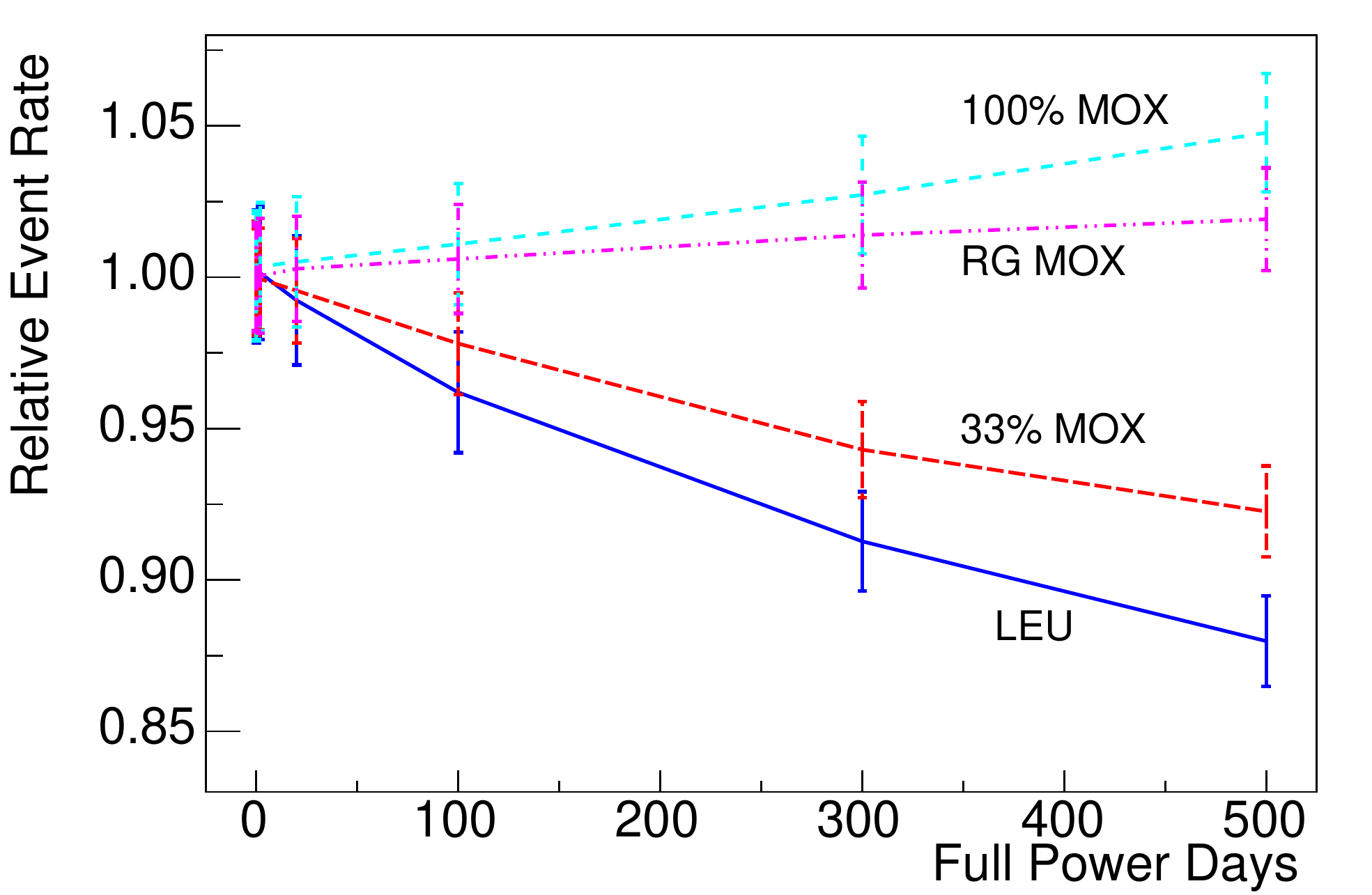}
\label{fig:sigReltoZero1MeV} 
%\subfigure[Antineutrino signal relative to LEU fuel.] {
%\includegraphics[width=0.48\textwidth]{./Figures/sigReltoLEU1MeV}
%\label{fig:sigReltoLEU1MeV}}
\end{center}
\caption{Detected antineutrino event rate for an antineutrino energy threshold of $3.3$~MeV, relative to the value at BOC with total uncertainties. }
\label{fig:WGMOX_RGMOX1MeV}
\end{figure}

\begin{comment}
\begin{figure}[h!]
\begin{center}
\subfigure[ Antineutrino rate relative to to zero burnup] {
\includegraphics[width=0.45\textwidth]{./Figures/sigReltoZero2MeV}
\label{fig:sigReltoZero2MeV} }
\subfigure[Antineutrino signal relative to LEU fuel.] {
\includegraphics[width=0.45\textwidth]{./Figures/sigReltoLEU2MeV}
\label{fig:sigReltoLEU2MeV}}
\end{center}
%\setlength{\abovecaptionskip}{-5pt}
\setlength{\belowcaptionskip}{0pt}
\caption{Relative antineutrino rate during one cycle (assumes 3.3 MeV threshold.)}
\label{fig:WGMOX_RGMOX2MeV}
\end{figure}

\begin{figure}[h!]
\begin{center}
\subfigure[ Antineutrino rate relative to to zero burnup] {
\includegraphics[width=0.48\textwidth]{./Figures/sigReltoZero5MeV}
\label{fig:sigReltoZero5MeV} }
\subfigure[Antineutrino signal relative to LEU fuel.] {
\includegraphics[width=0.48\textwidth]{./Figures/sigReltoLEU5MeV}
\label{fig:sigReltoLEU5MeV}}
\end{center}
%\setlength{\abovecaptionskip}{-5pt}
\setlength{\belowcaptionskip}{0pt}
\caption{Relative antineutrino rate during one cycle (assumes 5.8 MeV threshold.)}
\label{fig:WGMOX_RGMOX5MeV}
\end{figure}
\end{comment}

For either of the detector energy thresholds we have chosen, we will show in the following section that the different fuel loadings can be distinguished  via a statistical test, in which the measured rates are compared with a prediction based on operator declarations.
%\textbf{what does this mean -distinguish how?? - Adam  - clarify }. 
This is an important conclusion because a threshold-based background rejection strategy is significantly easier to implement than other methods of background suppression, such as the addition of shielding. Such an approach, involving measuring a raet above a well-defined energy threshold, simplifies deployments for this kind of monitoring task compared to a spectrum-based analysis, thereby increasing the possible range of applicability of antineutrino-based methods.   

As explained above, only the combined effects of power and burnup are constrained by the measured antineutrino rate evolution. This leaves any rate-based method susceptible to spoofing as long as the power is reported by the operator and not measured. For example, the reported power could be falsified over the entire cycle, in an attempt to mask the replacement of one core type with another. However, as we show here, a modest refinement of the present method can address the simplest version of this spoofing strategy. 

Figure~\ref{fig:WGMOX_RGMOX1MeV} shows antineutrino signal rate evolution for WGMOX, RGMOX, LEU and 1/3 WGMOX + 2/3 LEU cases normalized to the zero burnup value of the signal rate, for the $1.8$~MeV threshold case. As we show in the following section, the various fuel loadings can be distinguished via a statistical test, even in the presence of a constant shift in the operator-reported power. It is still possible for the operator to report a {\it varying} power level in order to mask the effect of a change in fuel type. However, this reported variation would itself be suspicious in a monitoring context, and is thus not an effective strategy. 

A few important trends are apparent in  Figure~\ref{fig:WGMOX_RGMOX1MeV}. As expected, the LEU signal decreases significantly with burnup. This is due to increased production of and reliance on $^{239}$Pu which emits fewer antineutrinos per fission above the inverse-beta detection threshold.  
%(This can also be determined from  Figures~\ref{fig:FR_Pu239} 
%and~\ref{fig:FR_Pu241}). 
On the other hand, the detected antineutrino rate is seen to decrease less rapidly, and then increase, as the fraction of MOX fuel increases. This is due to higher fission fraction provided by $^{241}$Pu, whose specific antineutrino emission rate is comparable to that of $^{235}$U.  Finally, a 100\% reactor-grade MOX core is seen to have a slightly higher absolute rate throughout the cycle compared with a 100\% weapons-grade MOX core, again due to the higher $^{241}$Pu fraction in the reactor-grade core. 
As we show in the next section, a hypothesis test can be used to take advantage of the observed differences in antineutrino rate evolutions, in order to make inferences about the fuel loading and fissile isotopic content of the core.
 
\section{Hypothesis Test for Anomalous Fuel Loading}
\label{sec:hyptest}
In this section we use a non-parametric hypothesis test, the Wilcoxon Ranked Sign Test (RST)~\cite{RST}, to determine whether different evolutions can be distinguished at a desired level of confidence. The ranked sign test is used to accept or reject the null hypothesis - in this case that the two distributions being compared are identical. The test compares a sequence of predicted and measured data values and calculates a \textit{p-value} for the data ensemble. If, under repeated applications of the test, the p-value is consistently below (above) a specific value  - here 0.05  - the null hypothesis is rejected (confirmed) with a significance corresponding to the chosen value.  Consistency of the measured and predicted antineutrino rate evolutions via the test can be used to infer that the fuel fissile isotopics and the reactor operations were consistent with declarations made by the party being monitored. 

\begin{figure}[ht!]
\begin{center}
\includegraphics[width=0.48\textwidth]{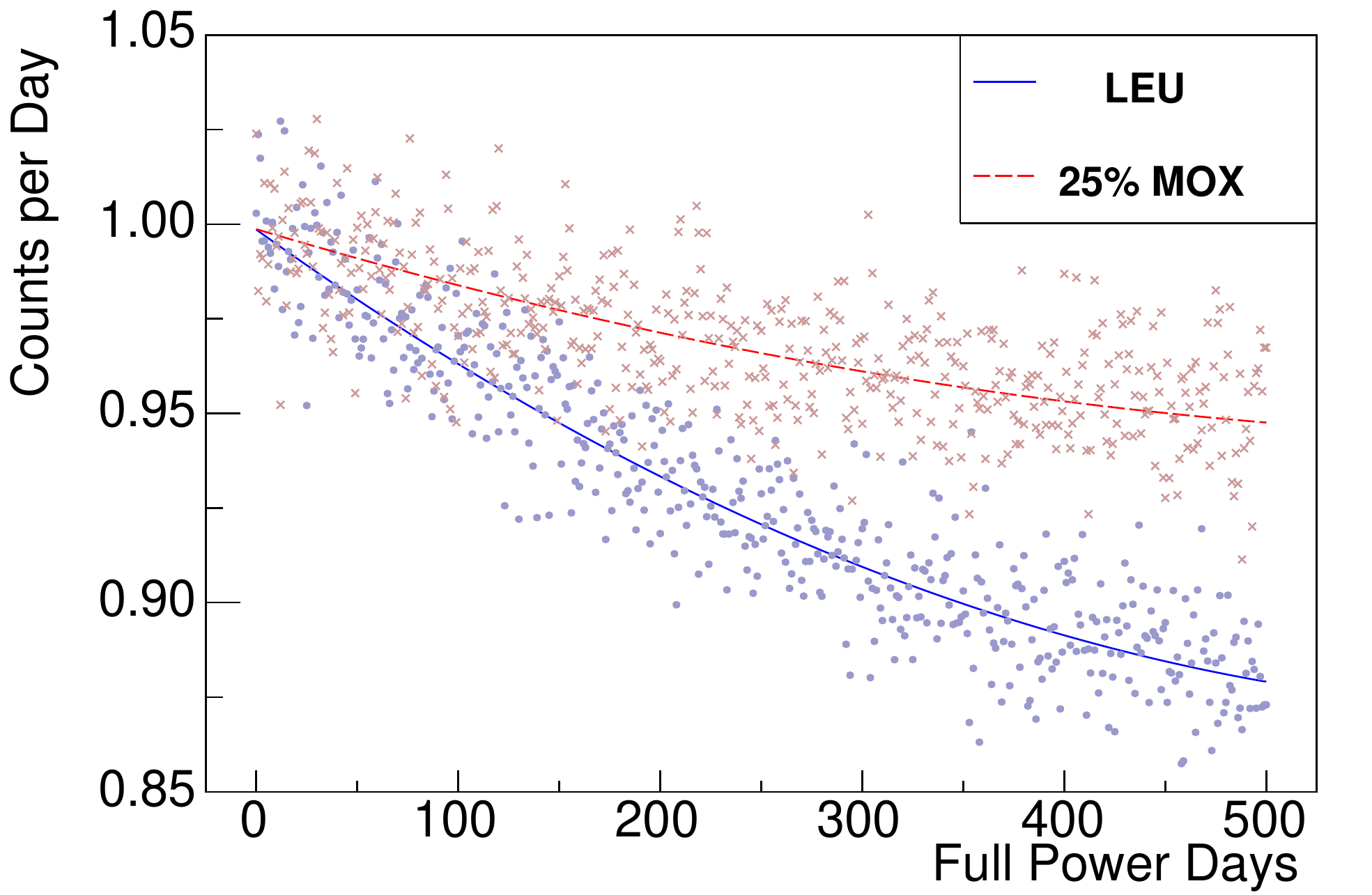}

\end{center}
\caption{A comparison of predicted and measured daily counts for a full LEU core and a 25\% MOX core.}
\label{fig:comparisonfig} 
\end{figure}

For the present test, we use the relative rates with errors as described in the previous section, assuming the 1.8~MeV detection threshold for the antineutrino energy. We assumed a constant, Gaussian distributed background equal to 25\% of the  detected antineutrino rate averaged over all evolutions. 
The treatment of errors for the  measured data set (here simulated) were summarized in the previous section. For the predicted data set (also simulated) we use the mean spectral and fission yield values as the most accurate estimate of these quantities.  The simulated measured and predicted data sets both account for the random variations arising from counting statistics, including the uncertainty introduced by the presence of background.     
Figure~\ref{fig:comparisonfig} shows an example evolution for a pair of predicted and measured data sets, for the case of 0\% MOX (LEU) and 25\% MOX cores.  

The RST is efficient at identifying systematic shifts in one population mean relative to another, even in the presence of significant correlated or uncorrelated uncertainty.  
Table~\ref{RSTtab} shows the number of days of data acquisition required to reach a \textit{p-value} that is less than $0.05$ for at least 90\% of all trials. We compare neighboring levels of MOX loading, as well as full weapons-grade and reactor-grade MOX cores. As a check on the method, we also examine the \textit{p-value} for 100 trials comparing identical 100\% WGPu MOX loadings. As expected the average significance for identical evolutions is uniform on the interval [0,1], with an average result of $p \simeq 0.5$.
 We take 0.05 to be the  \textit{p-value} threshold below which the null hypothesis is considered rejected, meaning the evolutions can be distinguished.  By this standard, all adjacent MOX loadings can be distinguished well before the end of a single 500-day cycle, with the exception of $75\%$ and $100\%$ WGPU MOX core loadings. To fix  the number of days required to achieve a given level of rejection,  we demanded that at least 90 of 100 ranked sign test trials, each with  measured and predicted data sets randomly generated according to the underlying distributions, gave a result of $p<0.05$.  Moreover, as seen in Table \ref{RSTtab}, a change in detection threshold from $1.8$~MeV  to $3.3$~MeV has only a modest effect on the number of days required to acheive the desired sensitivity.

\begin{figure}[ht!]
\begin{center}
\includegraphics[width=0.48\textwidth]{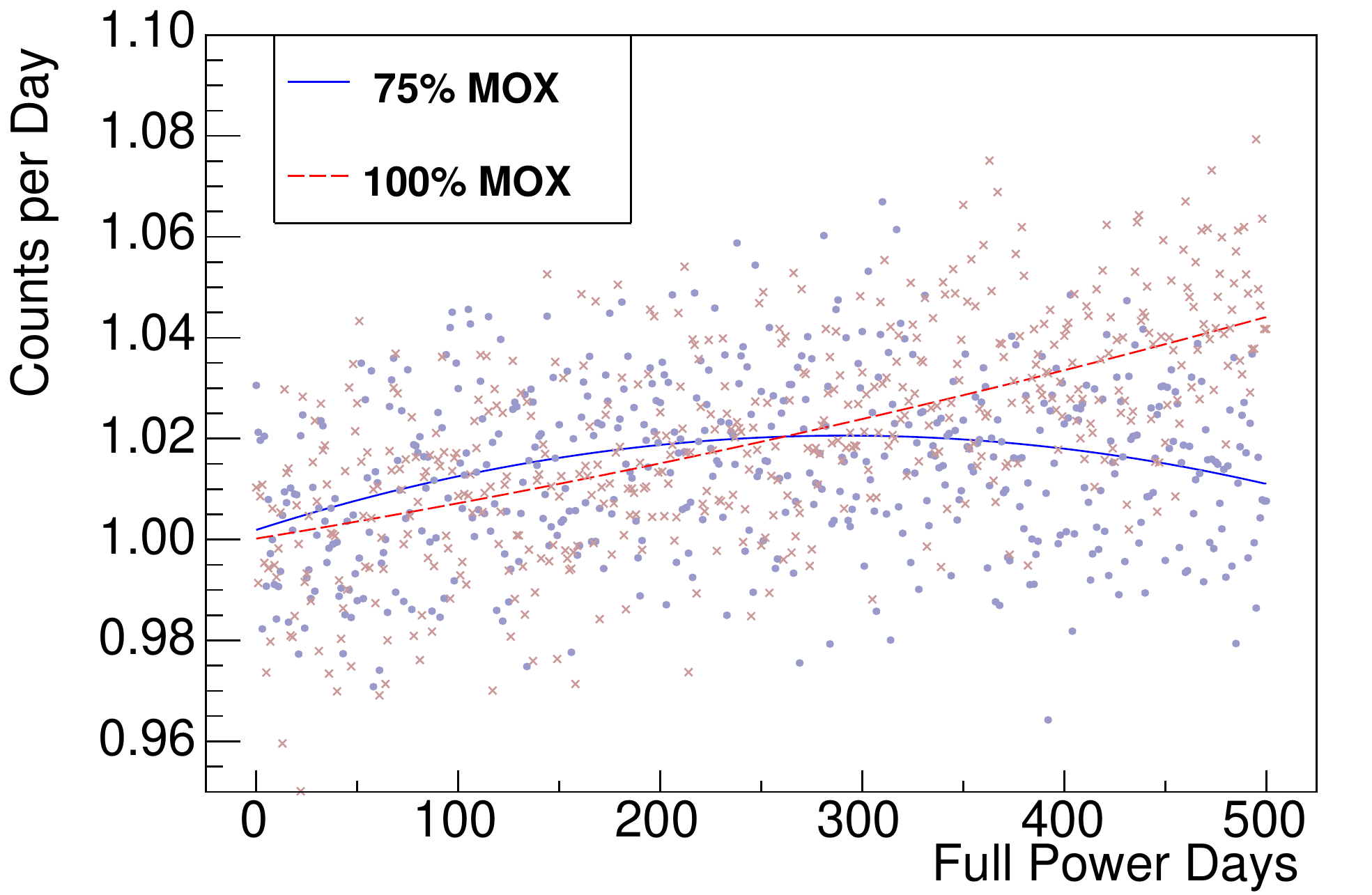}

\end{center}
\caption{A comparison of predicted and measured daily counts for a  75\% and 100\% MOX cores}
\label{fig:MOX100and75} 
\end{figure}

A visual comparison of the predicted $75\%$ and simulated measured $100\%$ evolutions, including all uncertainties, is shown in Figure~\ref{fig:MOX100and75}. The $75\%$ MOX evolution lies both above and below the $100\%$ evolution for different parts of the cycle. Because the ranked sign test is most sensitive to shifts in one direction, the test is poorly suited to this particular case.
Importantly, the results show that weapon-grade and reactor-grade MOX cores can be distinguished with high confidence within a single cycle. This discrimination capability is particularly important for plutonium disposition campaigns, for which the monitoring party has a strong interest in confirming that weapons-grade fissile materials are being dispositioned.

\begin{table}
\caption{ The approximate minimum number of days required to ensure an average  \textit{p-value} less than $0.05$ for 90 of 100 trials. Except for the comparison of predicted $75\%$ and  measured $100\%$ evolutions, all other evolutions are capable of being distinguished within a single 500 day evolution.}
\label{RSTtab}
\begin{tabular}{|p{2.0 cm}|p{2.0 cm}|p{1.5 cm}|p{1.5 cm}|}
 \hline
 Predicted MOX fraction & Measured MOX fraction & \multicolumn{2}{p{3 cm}|}{Minimum days required to achieve $\textit{p-value} <0.05$ for 90\% of RST trials, for the indicated detection threshold}  \\
 \hline
  \multicolumn{2}{|p{4.0 cm}|}{ } & $1.8$~MeV &$3.3$~MeV\\
        \hline
0\% 	& 25\%		&  80	    & 80 \\
25\%	& 33\%		&  150		& 165 \\
33\%	& 50\%		&  195		& 195 \\
50\%	& 75\%		&  54		& 60  \\
100\% WG& 100\% RG	& 300		& 315  \\
\hline
\end{tabular}
\end{table}

\section{Summary and Conclusions}

Antineutrino detectors offer an important capability that may be useful for a range of  modern safeguards and reactor monitoring applications. These detectors provide continuous knowledge about reactor operations. If core power and/or initial fuel loading are known, rate-based antineutrino detectors have a capability of deducing other operating parameters, in particular checking whether operator-declared core loading matches that of the actual operating core. The technology is well-established by multiple fundamental science and safeguards demonstrations. Because the detector can be located well outside the reactor core and requires to connection to plant systems, the installation is non-intrusive to reactor operations.  

Using assembly level simulations, we have shown that antineutrino flux decreases with burnup (or equivalently, cycle day at full power) for LEU cores, increases for most MOX fuel loadings, but does not appreciably change for cores with MOX fraction of about 75\%.  We also demonstrated that a hypothesis test that evaluates antineutrino rate information only can be used to distinguish a range of MOX loadings within a single cycle, including between WGPu and RGPu cores. For nonproliferation applications such as plutonium disposition, rate-based detectors offer important advantages in terms of simplicity of construction and operation, which likely facilitates their adoption in monitoring regimes.  

This assembly-level core simulation has more detail than is found in point-like core simulations.  Still more insight can be gained by simulating a more realistic mix of assemblies with various enrichment and burnup values in a single evolution. 

%\textbf{make some reference to spectrum? and RST section}

%The effect of these cycle-to-cycle variations in loading on sensitivity to MOX inventories are examined in a forthcoming companion paper. 

\begin{acknowledgments}
LLNL-JRNL-712137. This work performed under the auspices of the U.S. Department of Energy by Lawrence Livermore National Laboratory under Contract DE-AC52-07NA27344.
\end{acknowledgments}

\section{References}

\begin{comment}
\bibliographystyle{plain}

\end{comment}

%\nocite{*}
\bibliography{aipsamp}% Produces the bibliography via BibTeX.

\end{document}